\documentclass[a4paper,10pt]{article}
\usepackage{amsfonts}

\usepackage{makeidx}
\usepackage{amsmath}
\usepackage{amstext}
\usepackage{amsthm}
\usepackage{xspace}
\usepackage[dvips]{graphicx}

\newtheorem{theorem*}{Theorem}
\newtheorem{prop*} {Proposition}
\newtheorem{lemma*}{Lemma}

\theoremstyle{definition}
\newtheorem{definition}{Definition}
\newtheorem{definition*}{Definition}

\newtheorem{cor*}{Corollary}

\newtheorem{rem*}{Remark}
\theoremstyle{remark}

\newtheorem{dim*}{\bf Proof}

\newtheorem{guess*}{\bf Osservazione}

\newcommand{\no}{\noindent}

\input psfig.sty

\date{}

\begin{document}

\title{Data compression and genomes: a two dimensional life domain map}
\author{Giulia Menconi$^{a,b}$, Vieri Benci$^{a,b}$ and Marcello Buiatti$^{b,c}$\\
{\small {$^a$ Dipartimento di Matematica Applicata, Universit\`a\xspace di
Pisa}}\\
{\small {$^b$ C.I.S.S.C. Centro Interdisciplinare per lo Studio dei Sistemi
Complessi}}\\
{\small {Universit\`a\xspace di Pisa}}\\
{\small {$^c$ Dipartimento di Biologia Animale e Genetica, Universit\`a
\xspace di Firenze}}}

\maketitle

\begin{abstract}

We define the complexity of DNA sequences as the information content
per nucleotide, calculated by means of some Lempel-Ziv data
compression algorithm. It is possible to use the statistics of the
complexity values of the functional regions of dif\-fe\-rent complete
genomes to distinguish among genomes of different domains of life
(Archaea, Bacteria and Eukarya). We shall focus on the distribution
function of the complexity of noncoding regions. We show that the
three domains may be plotted in separate regions within the
two-dimensional space where the axes are the skewness coef\-ficient
and the curtosis coefficient of the aforementioned
distribution. Preliminary results on 15 genomes are introduced.
\end{abstract}
al\-go\-rithms, evolutionary dynamics\\[10pt]
\textbf{Corresponding author:} Dr. Giulia Menconi,
menconi@mail.dm.unipi.it\\Postal address: Dipartimento di Matematica
Applicata, Universit\`a\xspace di Pisa\\ Via Buonarroti 1C $-$ 56127
PISA ITALY.\\Telephone: +390502213879 $-$ fax: +390502213802
%%%%%%%%%%%%%%%%%%%%%%%%%%%%%%%%%%%%

\section{Introduction}

As discussed at length in an earlier paper \cite{buiattivarii}, genome
 evolution is characterised by some very peculiar features deriving
 from the structure and dynamics of living systems. Life in general is
 influenced at the same time by the variability generated by internal
 and external sources and by constraints deriving from
 self-organisation and adaptation dynamical rules. The interaction
 between these contrasting processes leads to generalised scale
 invariant behaviours of many living processes as reviewed by
 \cite{buiattivarii} and \cite{bmarco}. Genomes may be considered as
 the heritable ``data-base'' of living systems histories and are
 therefore a useful tool for the study of the dynamical features of
 the ratio between randomness and constraints throughout
 evolution. This is particularly relevant in view of the astounding
 changes in the conceptual framework of evolutionary theories deriving
 from the new data obtained through genome, transcriptome and proteome
 analyses in different classes of organisms from prokaryotes to
 eukaryotes. Evidence has been recently obtained showing that
 prokaryotes, eukaryotes and humans differ particularly as far as the
 sources of variability and homeorrhetic \cite{wad} processes are
 concerned. Prokaryotes, due to their short life cycles and to the
 haploid nature of their genomes, rely on the induction of genetic
 va\-ria\-bi\-li\-ty, through mutations, genome re-arrangements, DNA
 exchange within and between species; eukaryotes have developed
 throughout evolution very sophisticated and efficient processes
 leading to phenotypic plasticity; humans, the most ``generalist''
 specie in our Biosphere, adapt themselves through the exploitation of
 cultural variability. As a consequence of these different behaviours,
 prokaryote genomes are much smaller and contain a lower amount of
 regulatory non-coding sequences ($12\%$ in bacterium $E.\ Coli$) and
 a correspondingly higher level of coding ones, noncoding presence in
 multicellular eukaryotes ranging from $76\%$ in $Coenorhabditis$ to
 $98\%$ in humans \cite{shaba}. Due to the astounding ac\-ce\-le\-ra\-tion of
 whole genome sequencing, the statistical properties of DNA have been
 stu\-died extensively in the last fifteen years. The data obtained show
 that DNA is characterised by short and long range correlations linked
 to the functional role of different classes of sequences. Coding
 strings show -especially in eukaryotes- weaker long range
 correlations than non coding ones ( see for instance
 \cite{provata,goldberger,wentian,likaneko,voss}) and in general
 deviations from randomness are stronger in eukaryotes than in
 prokaryotes. Particularly, the authors of \cite{buiacqui2} used a
 simple model called ``Copying mistake map'', based on the
 superposition of a pure random choice model and an intermittent
 generator of homogeneous sequences with the aim of obtaining
 quantitative estimates of correlation intensities. The results
 obtained showed that the exponent of the power law was constant in
 all organisms while the fraction of long correlations increased in
 eukaryotes. However, it should also be noted that probably all the
 mentioned methods employed to measure the extent of deviations from
 randomness in DNA sequences may underestimate the impact of short
 range correlations and of short, local, low-complexity sequences with
 possible functional relevance on the basis of specific ``hidden
 conformational codes'' \cite{trifo}. The aim of the present paper is
 to gain some additional information on the dynamics of the ratio
 between randomness and constraints and its putative correlation with
 function, obtaining whole genome pictures on the basis of complexity
 measures carried out on DNA fragments with different functions,
 namely coding and non coding (introns and intergenic) in Archaea,
 Bacteria, Eukarya.
\section{Compression algorithms and DNA sequences}
We shall analyse the genome sequences from the point of view of data
compression in order to obtain a domain classification of genomes by
means of their Information Content as symbol sequences. When
considered as a single strand sequence of nucleotides, the genomes are
interpreted as symbol strings of finite length, drawn by an
Information Source (Nature) that remains mainly unknown and emits
symbols taken from the alphabet of the four nucleotides $\{A,\ C,\ G,\
T\}$. Each genome identifies a living organism and we assume that it
may be considered as the unique realisation produced by the Source
relative to that organism. We recall that, intuitively, an Information
Source is a device emitting a sequence of symbols $\dots
x_{1}x_{2}x_{3}\dots $ where each $x_{i}$ is an element of a finite
alphabet. DNA sequences are special quaternary symbol sequences. As
discussed in the Introduction, the sequences belonging to a living
organism are expected to be nonrandom due to internal and selective
constraints. Therefore DNA sequences should be compressible, at least
locally.

In our approach to symbol sequences, the crucial notion is the
\textit{Information Content}. Let $\mathcal{A}^*$ be the set of finite
sequences whose symbols belong to a finite alphabet
$\mathcal{A}$. Given some sequence $\tau\in\mathcal{A}^*$, the meaning
of \textit{\ quantity of information} $I(s)$ contained in $s$ has the
following natural connotation:

\begin{center}
$I(\tau)$ \textit{is the length of the smallest binary message from which you
can reconstruct} $\tau$.
\end{center}

In his pioneering work, Shannon defined the quantity of information as a
statistical notion using the tools of probability theory (\cite{kin}). Thus
in Shannon framework, the quantity of information which is contained in a
string depends on its context. For example the string $^{\prime
}pane^{\prime }$ contains a certain information when it is considered as a
string coming from the English language. The same string $^{\prime
}pane^{\prime }$ contains much less Shannon information when it is
considered as a string coming from the Italian language because it is more
frequent in the Italian language (in Italian it means ''bread'' and, of
course, it is very frequent). Roughly speaking, the Shannon information of a
string is the absolute value of the logarithm of the probability of its
occurrence.

However, there are measures of information which depend intrinsically on the
string and not on its probability within a given context. We shall adopt this
point of view. An example of these measures of information is the
Algorithmic Information Content ($AIC$). We shall not formally define it (see 
\cite{kin} for rigorous definitions and properties). We limit ourselves to
give an intuitive idea which is very close to the formal definition. We can
consider a partial recursive function as a computer $C$ which takes a
program $p$ (namely a binary string) as an input, performs some computations
and gives a string $\tau=C(p)$, written in the given alphabet, as an output.
The $AIC$ of a string $\tau$ is defined as the length of the shortest binary
program $p$ which gives $\tau$ as its output, namely 
\begin{equation*}
I_{AIC}(\tau,C)=\min \{|p|:C(p)=\tau\}, 
\end{equation*}
where $|p|$ means the length in bit of the string which the program
$p$ consists of. A theorem due to A. N. Kolmogorov (\cite{kolmogorov})
implies that the Information Content ${AIC}$ of $\tau$ with respect to
$C$ depends only on $\tau$ up to a fixed constant, therefore its
asymptotic behaviour does not depend on the choice of $C$. The
shortest program $p$ which outputs the string $\tau$ is a sort of
optimal encoding of $\tau$. The information that is necessary to
reconstruct the string is contained in the program.  Unfortunately,
this coding procedure cannot be performed by any algorithm
(\cite{VBook}). That is, the Algorithmic Information Content is not
computable by any algorithm.

Our method is focused on another measure: the Information Content of a
finite string can also be defined by a lossless data compression algorithm $Z
$ (\cite{cleary}). This turns out to be a Computable Information Content
($CIC$). In reference \cite{licatone} quantitative relations among Shannon
entropy of the source, the AIC and the $CIC$ of sequences are provided.

We shall therefore investigate whether it is possible to use the $CIC$
of the functional regions of different genomes to distinguish among
genomes of different domains of life (Archaea, Bacteria and
Eukarya). If the $complexity$ of some DNA sequence is defined as the
information content $per$ nucleotide, then we shall focus on the
distribution function of the compelxity values relative to noncoding
regions. We shall also show that the three domains may be plotted in
separate areas within the two-dimensional space where the axes are
the skewness coefficient and the curtosis coefficient of the
aforementioned distribution.

In this paper we shall use information content to extract some
phylogenetic relationships among complete genomes and identify
life domains. Ideas from data compression are not new in genome
analysis. Classical studies in compression algorithms on DNA sequences
answer the question about the com\-pres\-si\-bi\-li\-ty of DNA with
the additional advantage of using compression techniques to capture
the properties of DNA, for instance to identify where some linguistic
characteristic structures (such as reverse complements and approximate
repeats) are located within genomes. Several special-purpose
compression algorithms for DNA sequences have been developed (for
instance, see http://www.ebi.ac.uk for an overview of current research
groups at the European BioInformatics Institute). Part of these
so-called DNA-oriented algorithms have been also used to study pattern
matching problems (e.g. \cite{rivalstandem}). Moreover, techniques
based on the explicit calculation of some information content have
been developed to solve gene-finding problems (\cite{chenchang},
\cite{mm06}) and to construct metric distances from which reliable
phylogenetic trees may be built (see for instance
\cite{chen},\cite{caglioti}, \cite{sayood}). In this context,
information content's growth rate is a way to approximate the entropy
of the sequence, which is a measure of the complexity of the
sequence. In this paper we discuss the application of a variant of
Lempel-Ziv compression schemes (\cite{lz77},\cite{lz78}). Some other
modified algorithms have been used in \cite{gusev} and \cite{orlov}
for calculating genetic sequence complexity. Of course, this list of
references can not be exhaustive.

The notion of complexity of a finite symbol sequence is controversial
and basically twofold: it may be defined by paying attention to the
regularity of sequences (as in the case of Information-based methods),
or - alternatively - by focusing on quantifying the randomness of the
same sequences, as in the case of correlation analysis
(\cite{wentian}). In particular, the long-standing interest in
understanding the meaning of short-range and long-range correlations
in nucleotidic sequences showed that they are bound to several
specific features of the DNA code.  Statistical analysis of the
spatial distribution of nucleotides confirmed the mutual relation
between functional requirements and DNA heterogeneity; the presence of
differences between coding and noncoding regions were also studied in
a phylogenetic setting, in order to relate it to a selection pressure
on the DNA structure (\cite{voss}, \cite{lioruffo}, \cite{likaneko},
\cite{goldberger}, \cite{herzeltrifo},
\cite{audit},\cite{buiacqui1},\cite{buiacqui2} and references
therein). Recently, also the opposite direction has been
investigated: patterns in the short-range statistical correlations in
DNA sequences have beeen shown to serve as evolutionary fingerprints
of eukaryotes \cite{dehnert}.

Our analysis is set as follows. 

The information content (CIC) of any functional region within complete
genomes shall be calculated by means of a modified Lempel-Ziv
compression algorithm called CASToRe. The complexity of the functional
region is the information content per nucleotide. We show that this
kind of complexity analysis of complete genomes is an indicator of
$biological$ complexity of the organisms, but it is not subtle enough
to allow an evolution arrow to be rigorously defined. Furthermore, the
linguistic differences among genomes are {\it per se} sufficient to
allow a precise positioning of the genomes on the two-dimensional map
to be settled. Understanding the reasons of his result may be the
starting point of a novel way to apply compression algorithms to DNA
sequences. In particular, the applied algorithm and Lempel-ziv
schemes in general provide a final dictionary of recurrent words in
the analysed sequence, such paving the way to pattern identification.

Finally, as we shall clarify in section \ref{castore}, the choice of
CASToRe is motivated by the fact that this algorithm is more suitable
than LZ78 to identify regularities. Indeed, it is well-known that a
better compression may be reached by using specific DNA-oriented
algorithms. Nevertheless, here we are interested in the fact that also
a generic-purpose compression scheme may allow a life domain
classification to be achieved, such confirming the existence of
macroscopic features marking the genomes.
%%%%%%%%%%%%%%%%%%%%%%%%%%%%%%%%%%

\section{Computable Information Content}

Formally, a compression algorithm is a reversible coding such that
any original finite sequence $\tau$ over a finite alphabet $\mathcal{A}$ may
be recovered from the encoded string $Z(\tau)$.

\begin{definition}[Compression Algorithm]
A lossless data compression algorithm is any injective function
$Z:\mathcal{A}^{\ast }\rightarrow \{0,1\}^{\ast }$.
\end{definition}

Therefore, since the coded string $Z(\tau)$ contains all the information that is
necessary to reconstruct and describe the structural features of the
original string, we can consider the length of the coded string as an
approximate measure of the quantity of information that is contained
in the original string.

\begin{definition}[Computable Information Content]
The Information Content of a finite string $\tau\in\mathcal{A}^{\ast }$ with
respect to a compression algorithm $Z$ is defined as 
\begin{equation}
CIC_{Z}(\tau)=|Z(\tau)|\ .
\end{equation}
The $CIC$ of a string $\tau$ is the length (in bit units) of the coded string $Z(\tau)$.
\end{definition}

The advantage of using a compression algorithm lies in the fact that the
Information Content $CIC_{Z}\left( \tau\right) $ is a computable function over
the space of finite strings. For this reason we named it Computable
Information Content.

Moreover, we define another quantity, the complexity of a finite sequence,
providing an estimate for the rate of Information Content contained in it.

\begin{definition}[Complexity of a finite string]
The complexity of $\tau$ with respect to $Z$ is the information per symbol 
\begin{equation}
C_{Z}(\tau)=\frac{I_{Z}(\tau)}{|\tau|}\ .
\end{equation}
\end{definition}

The complexity value of sequence $\tau$ is related to its
compressibility in the sense that the lower is $C_{Z}(\tau)$, the more
compressible $\tau$ is.

\begin{rem*}
  Under suitable optimality assumptions on the compression algorithm
  $Z$, we can extend the definition of complexity to infinite symbolic sequences
  belonging to $\Omega _{\mathcal{A}}$ and asympotically obtain the
  Shannon entropy of the Information Source from which the sequence
  has been drawn (\cite{licatone}).  The theoretical work has been
  extended also to trajectories coming from general dynamical systems
  and it is supported by application to several complex systems, as to
  turbulent or intermittent regimes and to weakly chaotic dynamical
  systems (see \cite{licatone} and references therein).
\end{rem*}

\subsection{The algorithm CASToRe}\label{castore} 

We have created and implemented a particular compression
algorithm we called CASToRe which is a modification of the Lempel-Ziv
compression schemes $LZ77$ \cite{lz77} and $LZ78$ \cite{lz78} and it has
been introduced and studied in references \cite{CSF02} and \cite{menconi}.

First, we describe the internal running of CASToRe. Then, we briefly
compare CASToRe to LZ77 and LZ78.

As any Ziv-Lempel schemes, the algorithm CASToRe is based on an
adaptive dictionary (\cite{cleary}). At the beginning of encoding
procedure, the dictionary contains only the alphabet. In order to
explain the main rules of the encoding, let us consider a step $h$
within the encoding process, when the dictionary already contains $h$
phrases $\{e_1,\dots,e_h\}$.

The new phrase is defined as a pair (\textit{prefix pointer},
\textit{suffix pointer}). The two pointers are referred to two (not
necessarily different) phrases $\rho_p$ and $\rho_s$ chosen among the
ones contained in the current dictionary as follows. First, the
algorithm reads the input stream starting from the current position of
the front end, looking for the longest phrase $\rho_p$ matching the
stream. Then, the algorithm looks for the longest phrase $\rho_s$ such
that the joint word $\rho_p+ \rho_s$ matches the stream. The new
phrase $e_{h+1}$ that shall be added to the dictionary is then
$e_{h+1}=\rho_p+ \rho_s$. The output file contains an ordered sequence
of the binary encoding of the pairs $(i_p,i_s)$ such that $i_p$ and
$i_s$ are the dictionary index numbers corresponding to the prefix
word $\rho _p$ and to the suffix word $\rho_s$, respectively. The pair
$(i_p,i_s)$ is referred to the new encoded phrase $e_{h+1}$ and has
its own index number $i_{h+1}$.

{\bf Example.} Consider the following sequence:
$$\tau= TCTATCTGATTTTCTCCTGGATC $$ The starting dictionary is
$\{A,C,G,T\}$. At the end of the compression, the dictionary contains
9 new words, built as follows.
$$
\begin{array}{ccc}
{\begin{array}{c}
dictionary\\
index
\end{array}}&(prefix,suffix)&word\\
0&-&\mbox{A}\\
1&-&\mbox{C}\\
2&-&\mbox{G}\\
3&-&\mbox{T}\\
4&(3,1)&\mbox{TC}\\
5&(3,0)&\mbox{TA}\\
6&(4,3)&\mbox{TCT}\\
7&(2,0)&\mbox{GA}\\
8&(3,3)&\mbox{TT}\\
9&(8,1)&\mbox{TTC}\\
10&(4,1)&\mbox{TCC}\\
11&(3,2)&\mbox{TG}\\
12&(7,4)&\mbox{GATC}\\
\end{array}
$$ 

One of the basic differences in the coding procedure is that the
algorithm $LZ77$ splits the input strings in overlapping phrases,
while the algorithm CASToRe (as well as $LZ78$) parses
the input string in non-overlapping phrases. Moreover, CASToRe differs
from $LZ78$ because the new phrase is a pair of two already parsed
phrases, while $LZ78$ couples one already parsed phrase and one symbol
from the alphabet.

The reason for the acronym CASToRe (meaning Compression Algorithm
Sensitive To Regularity) is that this scheme provides a sensitive
measure of information content in low entropy sequences (see
\cite{menconi}). For instance, consider the case where $\tau=AAA\cdots
A$ is a constant sequence of length $n$. The theory predicts that the
best possible Information Content for a constant sequence of length
$n$ is $AIC(\tau) =\log (n) + $constant. If the algorithm $Z$ is
CASToRe, the value of the Information Content is $I_Z(\tau)=4+2\log
(n+1)[\log (\log (n+1))-1]$. It may be shown that if the algorithm $Z$
is $LZ78$, then $I_Z(\tau)=const\ +\ n^{\frac 1 2}$.  So, we cannot
expect that $LZ78$ is able to distinguish a sequence whose Information
Content grows like $n^{\alpha}$ (with $\alpha < \frac 1 2$) from a
constant or periodic string. Furthermore, the running time of CASToRe
is also sensibly shorter than that of $LZ77$ (with infinite window),
then any implementation is more efficient. These are the main reasons
that motivate the choice of using CASToRe also for these numerical
experiments.

\begin{rem*}
In the following, the compression algorithm to which we refer for
numerical experiments is the algorithm CASToRe. Therefore, we shall
omit the subscript $Z$ everywhere.
\end{rem*}
\section{Fragment Analysis}

We have calculated the complexity values of coding and noncoding
regions within 14 complete genomes\footnote{The genomes have been
downloaded by means of the GenBank sequence libraries
http://www.ncbi.nlm.nih.gov/Genbank/index.html} of some Archaea,
Bacteria and Eukaryotes, together with chromosomes II and IV of
\textit{\ Arabidopsis thaliana}. The complete list is the following.

\noindent\underline{Archaea}:

\begin{enumerate}
\item  \textit{Methanococcus jannaschii}

\item  \textit{Archeoglobus fulgidus}

\item  \textit{Methanobacterium thermoautrophicum}

\item  \textit{Pyrococcus abyssi}
\end{enumerate}

\underline{Bacteria}:

\begin{enumerate}
\item  \textit{Aquifex aeolicus}

\item  \textit{Escherichia coli}

\item  \textit{Bacillus subtilis}

\item  \textit{Haemophylus influenzae}

\item  \textit{Mycoplasma genitalium}

\item  \textit{Rickettsia prowazekii}

\item  \textit{Thermotoga maritima}
\end{enumerate}

\underline{Eukarya}:

\begin{enumerate}
\item  \textit{Arabidopsis thaliana} (chr. II and IV)

\item  \textit{Saccharomyces cerevisiae}

\item  \textit{Caenorhabditis elegans}

\item  \textit{Drosophila melanogaster}
\end{enumerate}

In order to take into account the biological functional constraints actually
existing among the bases within the genome and to highlight new features of
coding and noncoding regions, we have exploited a \textit{fragment analysis}.

\begin{definition}[Fragment]
We say that any exon, intron or intergenic region is a functional
\textit{fragment} of the genome sequence, according to the prediction
as it has been identified via molecular analysis, biological databases
and statistical tools.
\end{definition}

We shall calculate the complexity
$C(f)$ of each fragment $f$ and we shall introduce the following indexes:

\begin{itemize}
\item \textbf{FC} is the acronym for fragment complexity $C$
calculated via $CIC$.

\item  \textbf{FC$_{ex}$/FC$_{in}$/FC$_{inter}$} is the FC only of
exon/intron/intergenic fragments, respectively.

\item  \textbf{AFC($\gamma $)} is the average fragment complexity, obtained
as the mean value of the complexity $FC(f)$ over all the fragments $f$, both
coding and noncoding, within genome $\gamma$: 
\begin{equation*}
AFC(\gamma )=<FC(f)>
\end{equation*}

\item  \textbf{AFC$_{ex}$/AFC$_{in}$/AFC$_{inter}$} is the AFC only of
exon/intron/intergenic fragments, respectively.
\end{itemize}

\subsection{Results}

\begin{table}[tbp]
\begin{center}
\begin{tabular}{|c||c|c||c|}
\hline
\textit{Genome} & \textit{AFC} & $AFC_{inter}$&$\rho (AFC)$ \\ \hline\hline
\textit{Methanococcus jannaschii} & 1.935 & 1.998&1.031 \\ \hline
\textit{Archeoglobus fulgidus} & 2.024 & 2.101&1.042 \\ \hline
\textit{Methanobacterium thermoautrophicum} & 2.042 & 2.132&1.042 \\ \hline
\textit{Pyrococcus abyssi} & 2.021 & 2.116 &1.042\\ \hline\hline
\textit{Aquifex aeolicus} & 2.003 & 2.081 &1.031\\ \hline
\textit{Escherichia coli} & 2.050 & 2.131 &1.042\\ \hline
\textit{Bacillus subtilis} & 2.035 & 2.110& 1.042\\ \hline
\textit{Haemophylus influenzae} & 1.998 & 2.062 &1.031\\ \hline
\textit{Mycoplasma genitalium} & 1.996 & 2.060 &1.031\\ \hline
\textit{Rickettsia prowazekii} & 1.909 & 1.946 &1.020\\ \hline
\textit{Thermotoga maritima} & 2.040 & 2.147& 1.053\\ \hline\hline
\textit{Arabidopsis thaliana} (chr. II and IV) & 2.038 & 1.903& 0.935\\ \hline
\textit{Saccharomyces cerevisiae} & 1.958 & 1.972 & 1.010 \\ \hline
\textit{Caenorhabditis elegans} & 2.056 & 1.960 &0.952\\ \hline
\textit{Drosophila melanogaster} & 2.051& 1.948&0.952\\\hline
\end{tabular}
\end{center}
\caption{\textit{Comparison of AFC vs. AFC - only intergenic regions
for the 14 genomes. The values of $\rho (AFC)$ are the ratios
$AFC_{inter}\slash{AFC}$.}}
\label{afcEinter}
\end{table}
We have compared the average fragment complexity of the above
collection of genomes. The results are shown in Table \ref{afcEinter}.

First of all, we have to point out that frequently the values of both
coding and noncoding fragment complexity and of the associated $AFC$
are slightly bigger than 2 bit per symbol and this is much more
evident in Prokaryotic noncoding regions (see their
$AFC_{inter}$). This is due to the presence of several regions that
are shorter than 200 bp. This short length causes a disadvantage in
the compression: the statistics of words over 4 symbols is very poor
and the compression is definitely worse than in case of long
sequences. Any Lempel-Ziv scheme looks for recurrent words and in the
case of poor statistics it is more likely to find many short words
than a few longer word, which leads to lower
compressibility. Nevertheless, length constraints do not bias the
possibility to develop a reliable method for coding sequences
identification based on how does the information content grow within
each fragment: in \cite{mm06}, it is shown that around 92\% of genes
were correctly predicted in Prokaryotic genomes. Applications to
Eukaryotes are in progress. Anyway, since we are not interested in
achieving the best compression, but a complete overview of the features of $FC$
distribution, we think that such an investigation should
comprehend fragments of any length.

It should also be stressed that AFC does not represent the average
com\-ple\-xity of the different genomes as the relative length of coding
and non-coding fragments is very different in eukaryotes and
prokaryotes in the sense that one fragment represents a different
percentage of the whole genome in the two domains. This is also why, at
a first glance to Table \ref{afcEinter}, the a\-ve\-ra\-ges
$AFC$ and $AFC_{inter}$ do not reveal any immediate meaningful
difference among the genomes. It is by comparing those values (we used
the ratio $\rho (AFC)=AFC_{inter}\slash{AFC} $) that an evidence for
the higher compressibility of Eukaryotic genomes is provided: the
lower value of $\rho (AFC)$ is determined by the stronger regularity
and greater extent of noncoding regions in Eukaryotes (see also
results in \cite{dehnert}).

Still, a clearer distinction between Prokaryotes and Eukaryotes should
be found in a more profound analysis of fragment complexity.
Therefore, we shall study the distribution of the fragment complexity
following the functional type and separately for each genome.

Figure \ref{distrProEu} shows some normalised distributions of the
$FC$ of fragments. Plots (a), (b) and (c) are relative to Prokaryotes
\textit{Archaeoglobus fulgidus} and \textit{Escherichia coli}: thick
solid line refer to coding fragments, while crossed-dashed lines refer
to noncoding fragments. Plots (d) and (e) are relative to Eukaryotes
\textit{Arabidopsis thaliana} and \textit{Saccharomyces cerevisiae}: thick
solid line refer to exons, crossed-dashed lines refer
to intergenic fragments and dashed lines refer to introns.

The fragment complexity $FC$ distribution appears to significantly
vary in coding and in noncoding regions: in the first case, the
distribution is generally more concentrated around the mean value,
while in the latter case the values of $FC$ are dispersed and the
oscillations have a smaller extent. Furthermore, especially in the
case of intergenic fragments, the distribution may present multiple
peaks and the decay from the maximum to zero (the so-called
\textit{tail} of the distribution) may be not as fast as in the case
of coding regions. Finally, as also Table \ref{afcEinter} suggests, in
Prokaryotes coding regions tend to have lower complexity than
noncoding regions, while this feature is inverted in
Eukaryotes. Again, this has two main motivations: regularities and
periodicities are stronger in noncoding regions of Eukaryotic genomes,
where moreover Prokaryotic noncoding fragments are shorter and less
correlated.

\begin{figure}
%\begin{tabular}{lr}
%(a)&(b)\\
%{\raggedright{\psfig{figure=globusnormok.ps,width=7cm,angle=270}}}
%&
%{\raggedleft{\psfig{figure=colinormok.ps,width=7cm,angle=270}}}
%\end{tabular} 
%\centerline{\psfig{figure=toganormok.ps,width=7cm,angle=270}(c)}
%\begin{tabular}{lr}
%{\raggedright{\psfig{figure=sacnormok.ps,width=7cm,angle=270}}}
%&
%{\raggedleft{\psfig{figure=arabnormok.ps,width=7cm,angle=270}}}\\
%(d)&(e)
%\end{tabular}
\caption{\it The normalised distribution of the Fragment Complexity $FC$ for 
  the different functional fragments: coding, intergenic and intron (when present).
  The pictures show the value of $FC$ vs. the frequency of fragments
  with that $FC$. In particular, only the subregion
  $\{FC\in[0.6,2.8]\} $ is drawn since the remaining part does not
  contain values different from zero. The plots are relative to (a)
  archaea Archaeoglobus fulgidus, bacteria (b) Escherichia coli and
  (c) Thermotoga maritima, (d) monocellular eukaryote Saccharomyces
  cerevisiae and (e) plant Arabidospis thaliana. Thick solid lines are
  relative to coding regions, while crossed-dashed lines are relative
  to intergenic regions and $-$ only in plots (d) and (e) $-$ dashed
  lines refer to introns.}
\label{distrProEu}
\end{figure}

The particular features that characterise the shape of the
distribution of $ FC$ in the intergenic regions of all the genomes led
us to the idea of studying some statistical indexes referred to the
distribution and measuring the ``spread'' of the distribution shape:
we focused on the \textit{curtosis coefficient} and on the
\textit{skewness coefficient}.

As we have previously pointed out, the distributions of $FC$ of intergenic
regions have long tails that are usually asymmetric with respect to the mean
value, so we may conclude that mean value and standard deviation do not
finely describe the core of the distribution. Furthermore, the degree of
convexity of the distribution curve may be a discriminant value to identify
the distribution. These two features (convexity and asymmetry) are
measured by the curtosis coefficient and by the skewness coefficient,
respectively, and they determine the ``distance in shape'' of the current
distribution with respect to a Gaussian distribution.

Let us define them. Let $X=(X_i)_{i\in I}$ be the finite data set of
the distribution of the fragment complexity $FC_{inter}$ of intergenic
fragments over the collection of the 14 genomes.

The \textit{standard deviation} $\sigma$ is, by definition, such that $
\sigma^2=\mathbb{E}[(X-\mathbb{E}[X])^2]$, where $\mathbb{E}[X]$ is the mean
value of the distribution, which we denoted by $AFC_{inter}$.

The \textit{curtosis coefficient} is calculated by means of the fourth
moment: 
\begin{equation*}
c=\frac{\mathbb{E}[(X-\mathbb{E}[X])^4]}{\sigma ^4}\ .
\end{equation*}
The higher the curtosis $c$ is, the flatter and more convex the distribution
is.

The \textit{skewness coefficient} is calculated by means of the third
moment: 
\begin{equation*}
s=\frac{\mathbb{E}[(X-\mathbb{E}[X])^3]}{\sigma ^3}\ .
\end{equation*}
The greater the skewness $s$ is, the more asymmetric the distribution is. If
the skewness is positive the asymmetry prevails on the left tail, otherwise
on the right tail.

\begin{figure}
%\centerline{\psfig{figure=disegnoAC.ps,width=12cm,angle=270}}
\caption{\it Statistical 2D-map for several genomes from the life
domains. The statistical coefficients Skewness and Curtosis are referred
to the distribution of only intergenic fragment
complexity. Prokaryotes: Stars ($\ast$)= Archaea; diagonal crosses
($\times$)= Bacteria. Eukaryotes: vertical crosses ($+$). Numbers and
letters are referred to the following genomes. Archaea: a)
Methanobacterium thermoautotrophicum, b) Archaeoglobus fulgidus, c)
Pyrococcus abyssi, d) Methanococcus jannaschii. Bacteria: 1)
Thermotoga maritima, 2) Aquifex aeolicus, 3) Haemophylus influenzae,
4) Mycoplasma genitalium, 5) Escherichia coli, 6) Bacillus subtilis,
7) Rickettsia prowazeckii. Eukaryotes: A) Saccharomyces cerevisiae, B)
Caenorhabditis elegans, C) Arabidopsis thaliana, D) Drosphila
melanogaster. }\label{AsiCur}
\vskip 0.7truecm
%\centerline{\psfig{figure=cfrdisegnoAC.ps,width=12cm,angle=270}}
\caption{\it Same picture as Figure \ref{AsiCur}, now highlighting the
boundaries among life domains. Stars ($\ast$): Archaea. Diagonal
crosses ($\times$): Bacteria. Vertical crosses ($+$):
Eukaryotes.}\label{sepAsiCur}
\end{figure}

Figure \ref{AsiCur} shows a two-dimensional picture that is a snapshot
in the space $ skewness\times curtosis$ where the values $(s,c)$ for
several genomes are plotted. The stars ($\ast$) are referred to the
Archaeal genomes, diagonal crosses ($ \times$) are relative to
Bacterial genomes, while vertical crosses ($+$) make reference to
Eukaryotic genomes. Numbers and letters on the picture identify the
genomes, as listed in the caption of Figure \ref{AsiCur}.

The three domains of life may be identified as three well separated
areas in the picture. The boundaries among Archaea, Bacteria and
Eukaryotes are plotted in Figure \ref{sepAsiCur}. 

Prokaryotes are strictly localised, almost collinear, at the bottom
horizontal strip. In fact, the flatness of their distributions is more
accentuated than in Eukaryotes, which are grouped at the top of the
picture. Therefore, a preliminary analysis suggests that the crucial
parameter in distinguishing between Prokaryotes (Archaea and Bacteria)
and Eukaryotes is the curtosis coefficient, which is significantly
higher in the Eukaryotes than in Prokaryotes. This coefficient may be
useful as a ``primitive'' evolutionary index.

For what concerns the separation between Archaea and Bacteria, even if
it is evident that the skewness coefficient plays a crucial role in
ordering the genomes, it is also clear that there is not a neat
boundary, because of the presence of a bacterium in half-strip that
contained all the results relative to the Archaeal genomes (i.e. the
genome labelled by 1 in the plots). This Bacterium, whose skewness and
curtosis coefficients are both the lowest $(s=0.180,\ c=0.137)$, is
\textit{Thermotoga maritima}. However, it should be stressed that it
was originally classified as an Archaea and now it is known as one of
the deepest and most slowly evolving lineages in thermophilic
Eubacteria. Presumably, this fact may motivate the tendency of the
distribution of the $FC$ of intergenic regions of \textit{Thermotoga
maritima} to become similar to an archeal distribution: Figure
\ref{distrProEu} shows that the similarity is not only qualitative
(multipeaked and widespread), but also quantitative. Moreover,
evidences have been showed (\cite{nelson}) for massive lateral gene
transfer between \textit{Thermotoga maritima} and bacterial genomes
(especially {\it Pyrococcus horikoshii}). Figure \ref{sepAsiCur}
supports the idea that also intergenic parts of \textit{T. maritima}
are Archaea-like.

\section{Concluding remarks}

The results previously discussed show that compression algorithms may
indeed be a useful tool for the study of the evolutionary dynamics of
the randomness/constraints ratio in genomes. The fragment complexity
is a measure of the relative impact of constraints on genome structure
and it differs from other traditional approaches based on long and
short-range correlations, periodicities, local DNA structures, etc
basically in two points. First, it is a self-contained measure and
gives a description of DNA sequences independently of the context they
belong to. Second, it is not directly connected to any statistical
quantity (for instance, the occurrence of some specific
oligonucloetides or other patterns), therefore may be used to compare
coding and non coding fragments even con\-si\-de\-ring that they have
different lengths in different domains. On the other hand through the
study of functionally different fragments , the presence of local low
or high complexity sub-sequences can be evaluated and eventually
correlated with its effects on DNA local conformational landscapes and
eventually with functional and structural differences between the
fragments studied \cite{buiacqui2}. In our case we show that
eukaryotic fragments are on the average more compressible than
prokaryotic ones coherently with the experimental finding of a higher
number of low complexity sequences particularly in non coding strings
of eukaryotes and with the vast literature on the effects on gene
expression of such strings (see for instance
\cite{buiacqui1,buiacqui2}). Moreover, the distribution of fragment
complexity values was found to be rather different in prokaryotes and
eukaryotes and to be a feature capable of discriminating between and
within the two groups of organisms. Particularly, it has been possible
to distinguish between Bacteria and Archaea through the skewness of
their distributions, Archaea curves being more near to a Gaussian than
those of Bacteria.  Moreover both Prokaryote groups differ from
Eukaryotes for the parameter curtosis whose values are higher in more
complex organisms than in then unicellular Saccharomyces. According to
\cite{allegr}, higher values of curtosis mean slow decay of the power
law in correlated systems. In other words we could infer from these
data that in eukaryotes there is a divergence between two groups of
sequences showing different complexity values. This is again coherent
with experimental data showing than in eukaryotes coding and non
coding sequences do indeed diverge in the number of low complexity
sequences while this kind of heterogeneity is much weaker in the case
of prokaryotes. Most probably, this is the result of selection forces
acting differently on prokaryotes and eukayotes: while in
prokaryotes'evolution selection acts mainly on genes (coding
sequences), in eukaryotes it is more effective on non coding regulatory DNA
(where adaptation relies, as discussed before, on regulatory
plasticity during life cycles). One striking example of this behaviour
is the fact that while genes -with very few exceptions- are very
similar in humans and schimpanzees, gene expression is differently
regulated. Different regulation therefore may have required the
fixation throughout evolution of more constraints in non coding
sequences and in general a wider divergence from coding ones.\\[10pt]
{\bf{Acknowledgement.}} The authors wish to thank Dr. Claudia Acquisti
(Center for Evolutionary Functional Genomics, Arizona State
University), for making genomic sequences available and easily
readable to symbolic analysis.

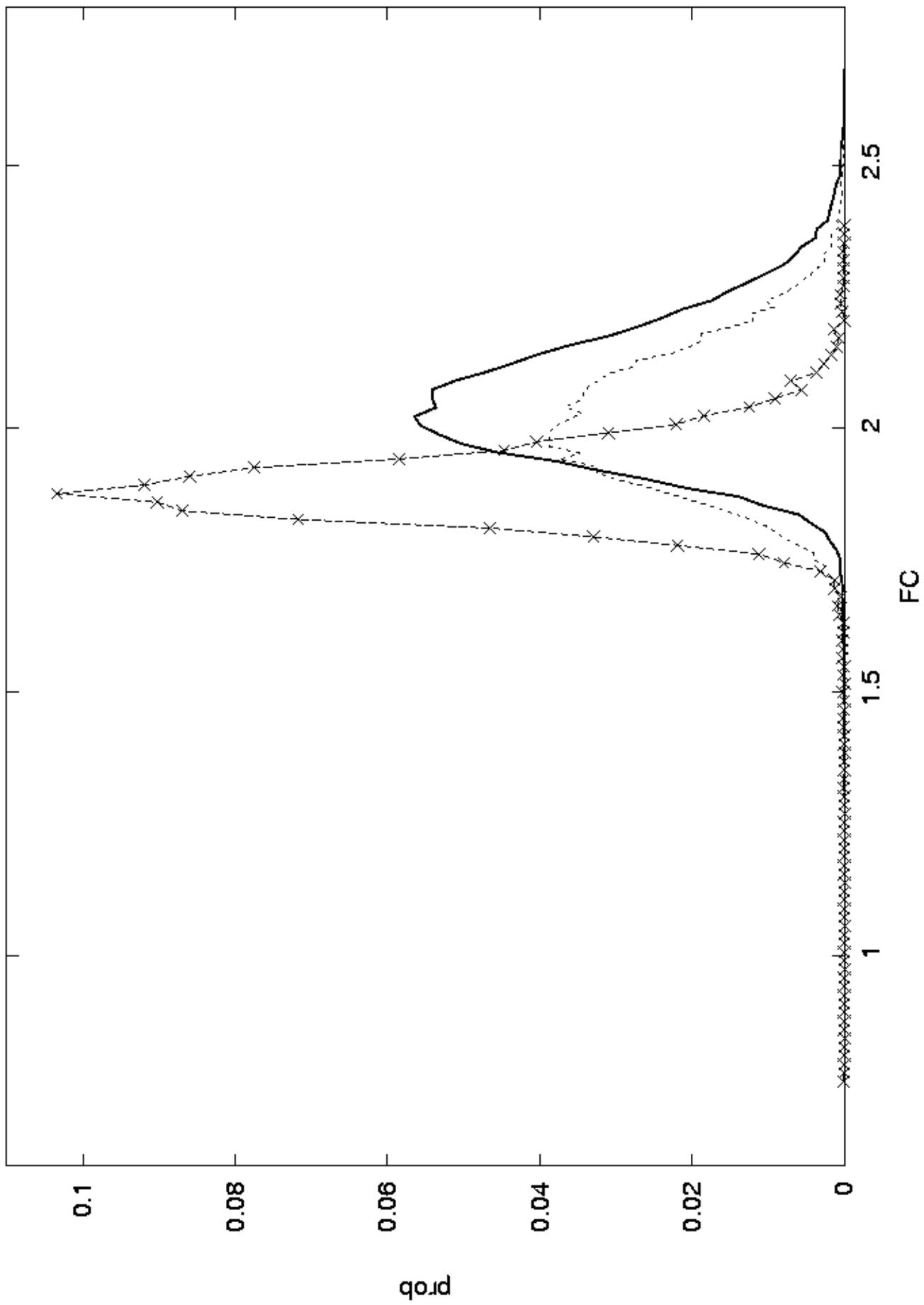
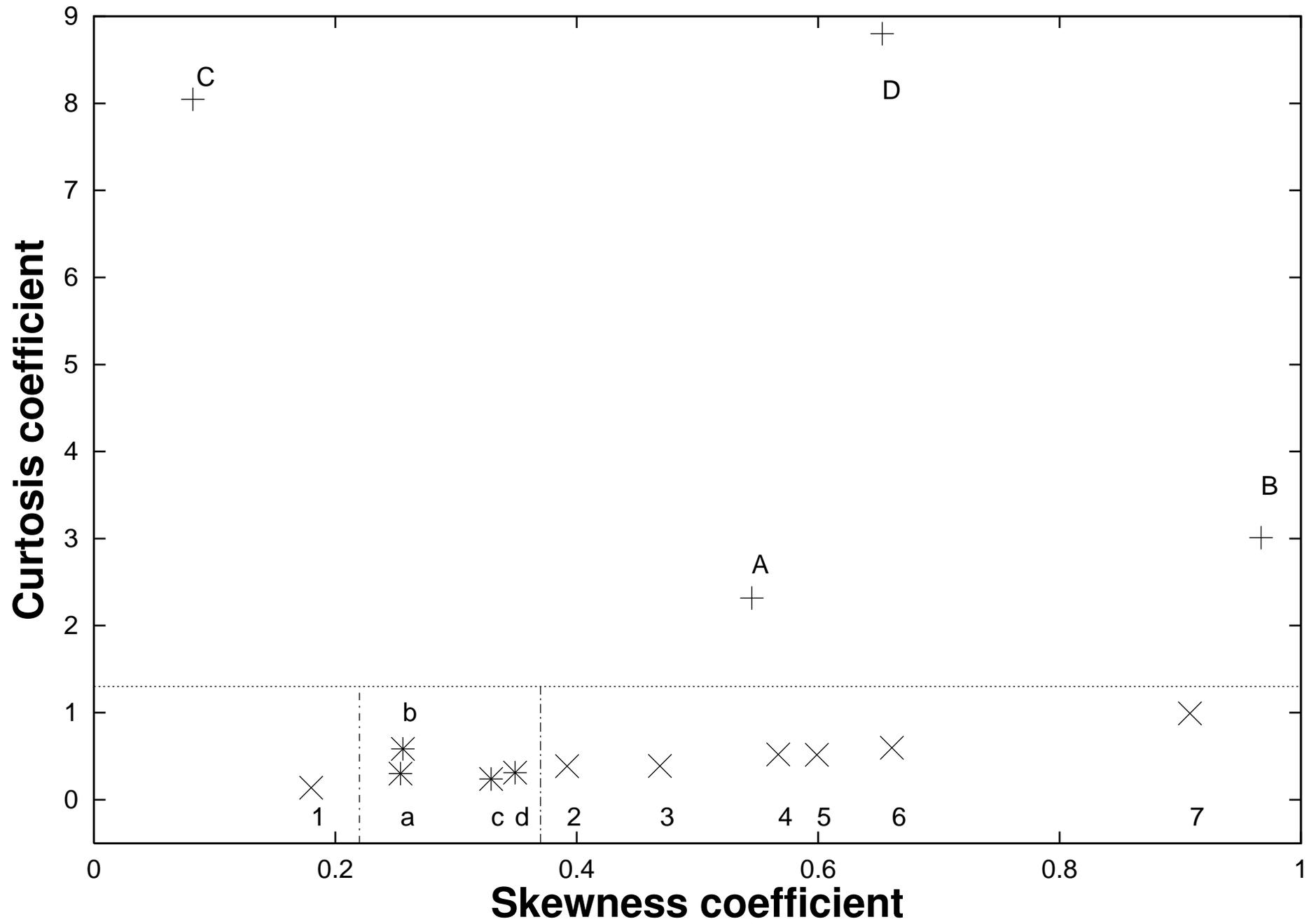
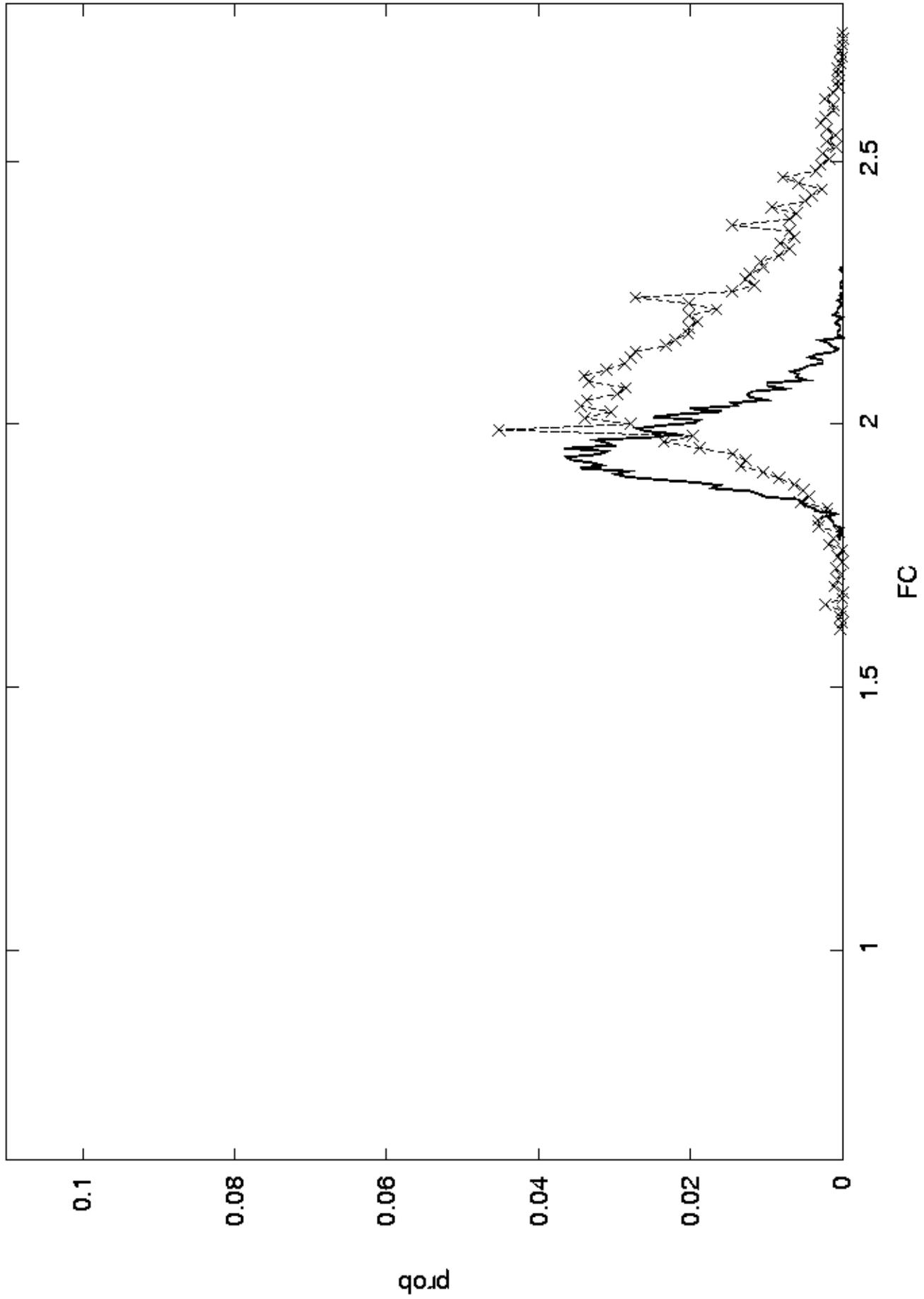
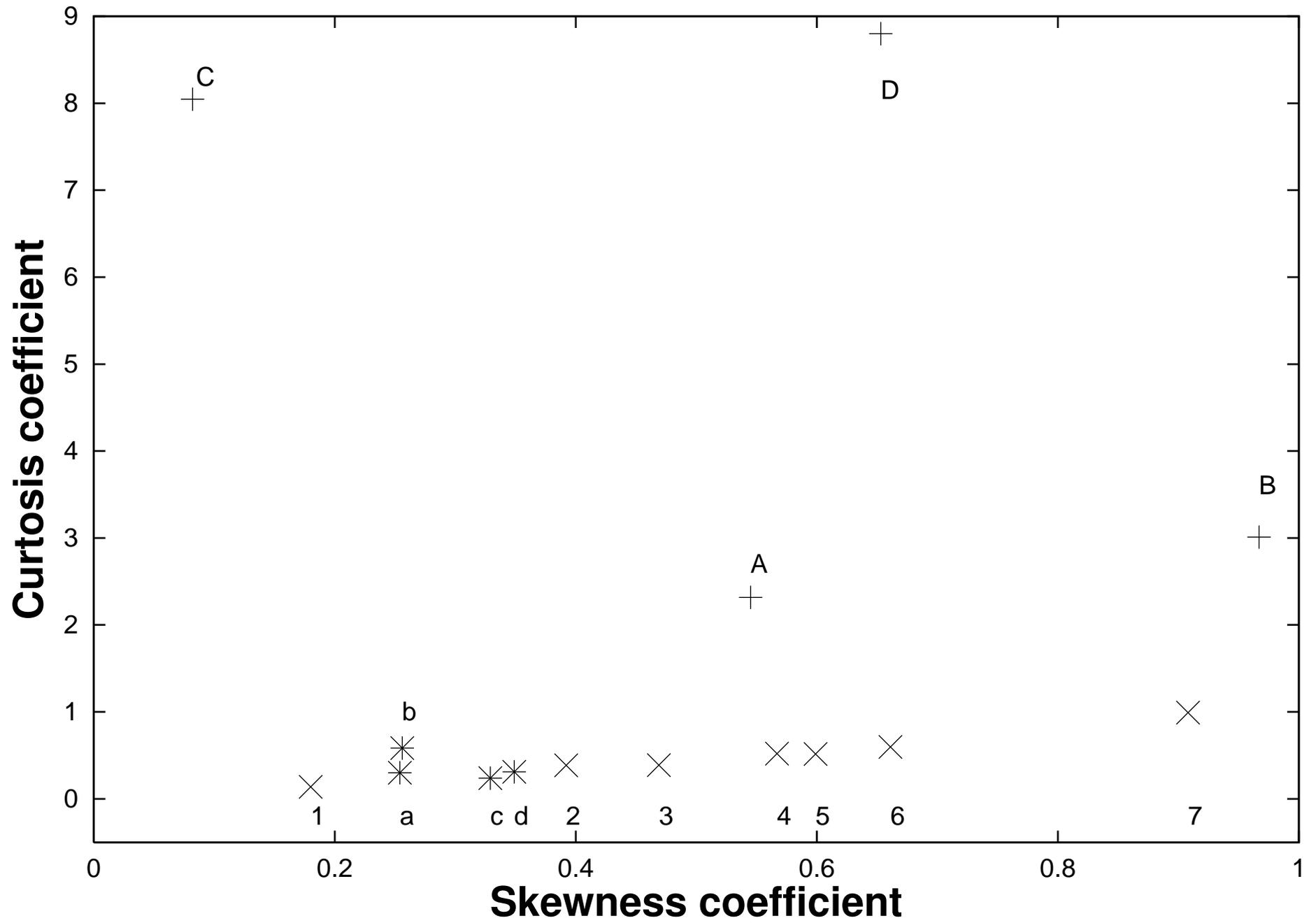
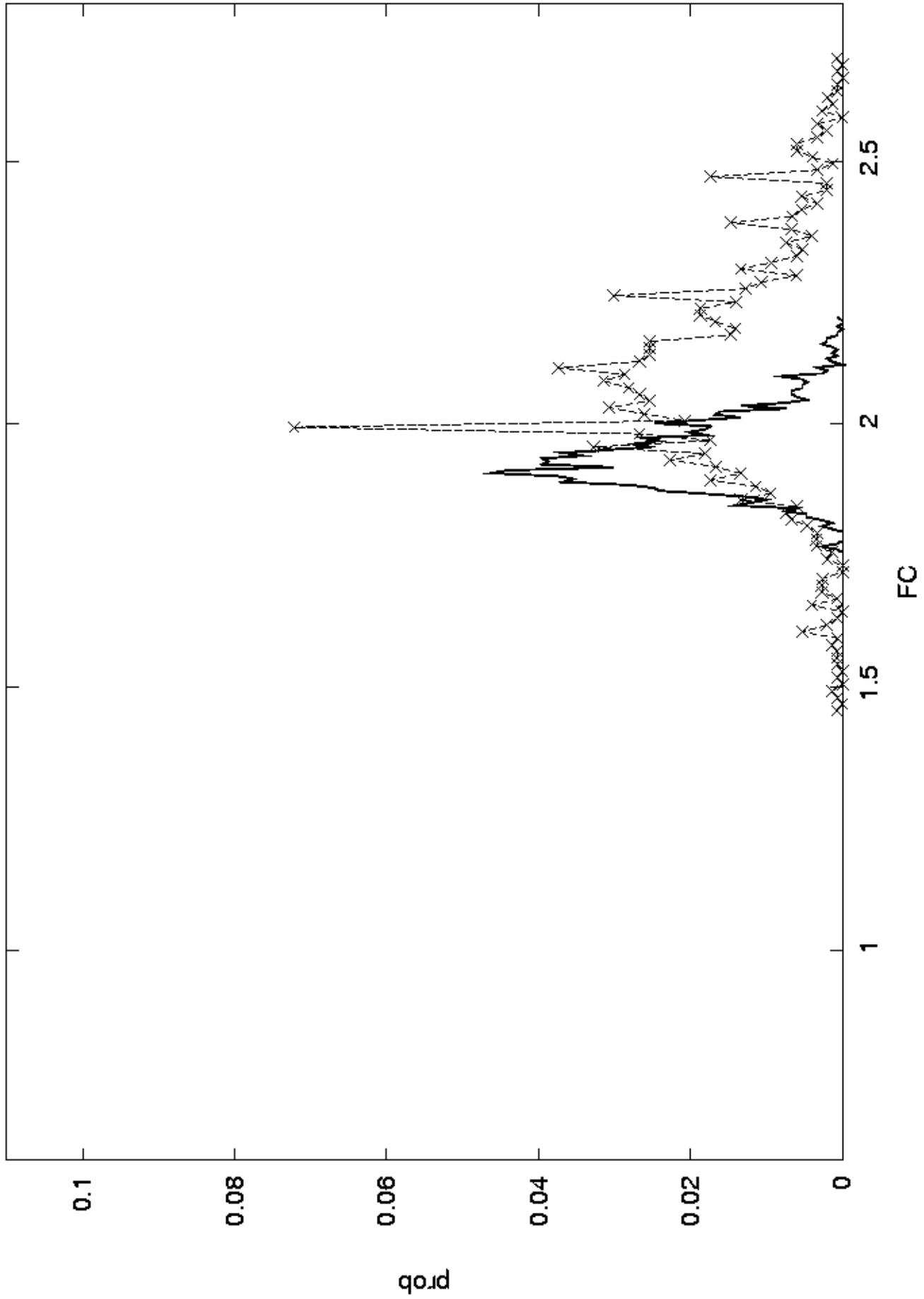
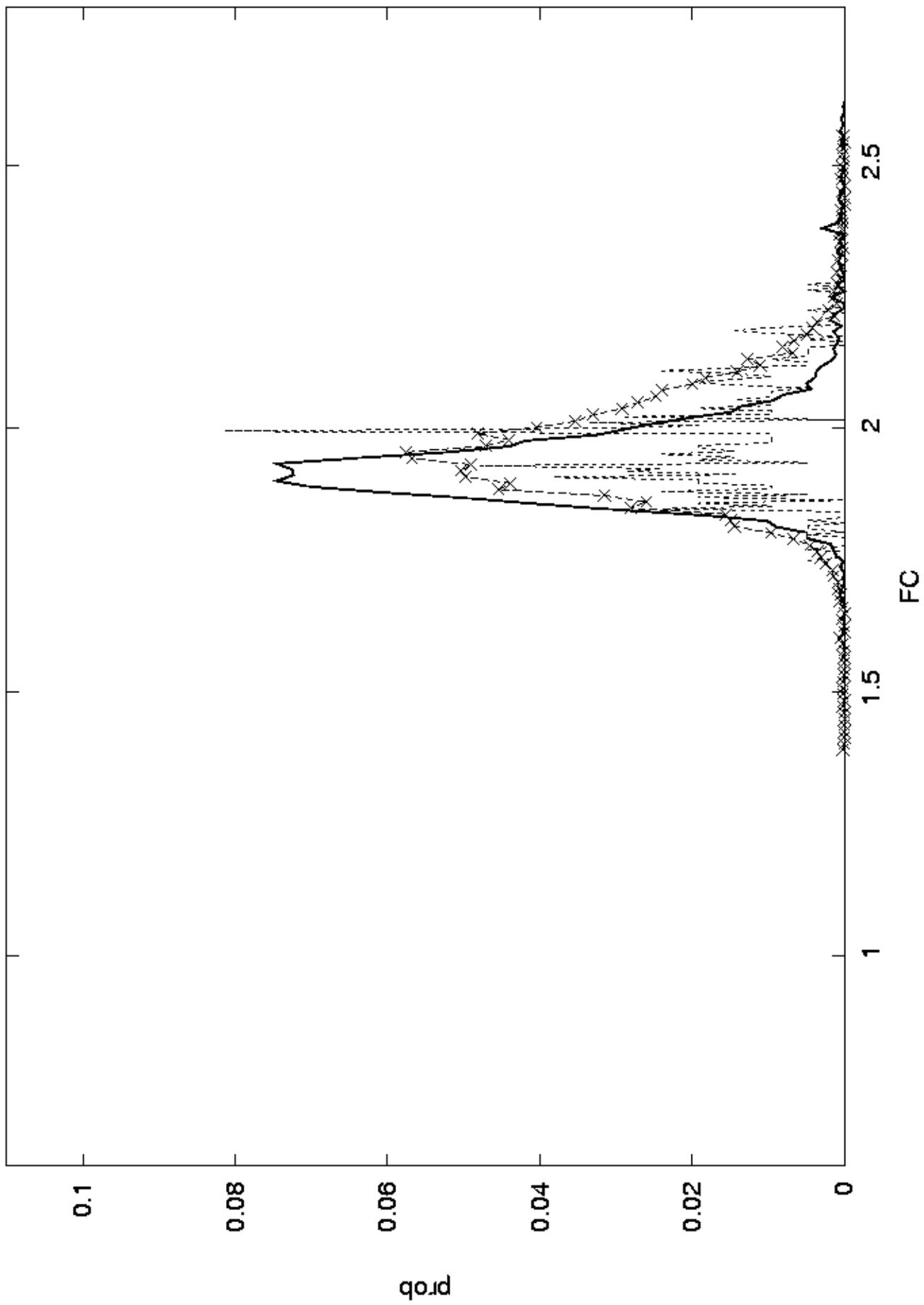
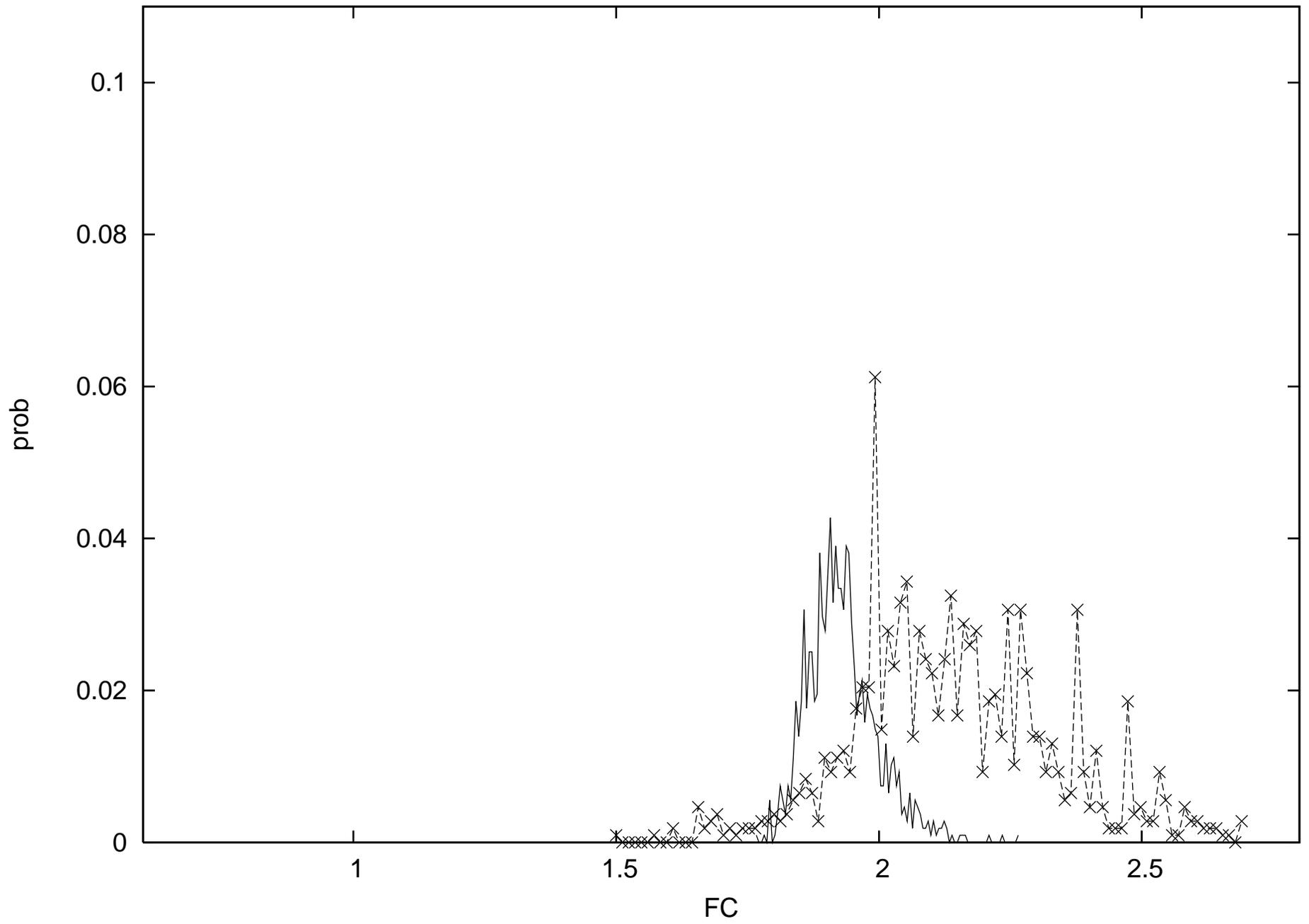
\end{document}